# Trion-Engineered Multimodal Transistors in Two-dimensional Bilayer Semiconductor Lateral Heterostructures

Baisali Kundu[1], Poulomi Chakrabarty,[2] Avijit Dhara[3], Roberto Rosati[4], Chandan Samanta[2], Suman K. Chakraborty[1], Srilagna Sahoo,[2] Sajal Dhara[3], Saroj P. Dash,[5] Ermin Malic[4*], Saurabh Lodha[2*], Prasana K. Sahoo[1*]

[1]*Quantum Materials and Device Research lab, Materials Science Centre, Indian Institute of Technology Kharagpur, West Bengal, India*
[2]*Department of Electrical Engineering, Indian Institute of Technology Bombay, Mumbai, India*
[3]*Department of Physics, Indian Institute of Technology Kharagpur, West Bengal, India*
[4]*Department of Physics, Philipps-University at Marburg, Marburg, Germany*
[5]*Department of Microtechnology and Nanoscience, Chalmers University of Technology, SE-41296, Göteborg, Sweden*
*\* prasana@matsc.iitkgp.ac.in; slodha@ee.iitb.ac.in; ermin.malic@uni-marburg.de*

## Abstract

Multimodal device operations are essential to advancing the on-chip integration of two-dimensional (2D) semiconductors in electronics, photonics, information and quantum technology. Precise control over carrier dynamics, particularly exciton generation and transport, is crucial for finetuning the functionality of optoelectronic devices based on 2D semiconductor heterostructure. However, the traditional exciton engineering methods in 2D semiconductors are mainly restricted to the artificially assembled vertical p-n heterostructures with electrical or strain-induced confinements. In this study, we utilized bilayer 2D lateral n-p-n multi-junction $MoSe_2$-$WSe_2$-$MoSe_2$ heterostructures with intrinsically spatially separated energy landscapes to achieve preferential exciton generation and manipulation without the need for external confinement. In lateral n-p-n field-effect transistor (FET) geometry, we uncover unique and non-trivial electro-optical properties, including dynamic tuning of channel photoresponsivity from positive to negative. The multimodal operation of these 2D FETs is achieved by carefully adjusting electrical bias and the impinging photon energy, enabling precise control over the trions generation and transport. Cryogenic photoluminescence measurement revealed the presence of trions in bilayer $MoSe_2$ and intrinsic trap states in $WSe_2$, which enhanced the sensitivity of the proposed device to near-infrared photons. Measurements in different FET device geometries show the multi-functionality of 2D lateral heterostructure phototransistors for efficient tuning and electrical manipulation of excitonic characteristics. Our findings pave the way for developing practical exciton-based transistors, sensors, multimodal optoelectronic on-chip and quantum technologies.

**Introduction**

Engineering light-induced charge carriers, such as excitons and trions, and their transport in two-dimensional (2D) semiconductor heterostructures devices based on transition metal dichalcogenides (TMDs) is crucial for versatile on-chip optoelectronics devices, neuromorphic computing, memory and quantum sensing and communication applications.[1–4] These 2D heterostructures possess vital properties such as controlled generation, transport, and confinement of carriers, strong light-matter interaction, ultra-fast speeds, transparency, flexibility, and the ability to be transferred deterministically on demand.[5–9] Such features make them ideal for on-chip integration onto electro-optical devices that rely on exciton-based signal processing.[1,10–12] Various device architectures have been investigated to manipulate charge carriers in 2D TMD heterostructures-based transistors. Of particular importance is the ability to dynamically tune and electrically control the channel conductivity on demand in 2D transistors. Additionally, on-chip electro-optical integration of these devices frequently requires external stimuli, such as photons of varying energy, to control and efficiently switch the channel conductivity. The challenges go beyond understanding light-induced anomalies in channel conductivity, extending to integrating 2D transistors into circuits for sensing and computing platforms, all within the constraints of the 2D "flatland". Addressing these challenges is essential to harness the potential of 2D heterostructure devices for future technologies.

Over time, several efforts have been made to manipulate carrier flow under light exposure in various 2D TMD heterostructures and other 2D devices, enhancing their photo-sensing characteristic typically through positive photoconductance (PPC).[13–15] Certain anonymous behaviours, such as light-induced negative photoconductivity (NPC), are often attributed to environmental factors like metastable traps, moisture absorption, deep-level excitation, and photo-bolometric effect. However, various proof of concepts have been explored to achieve negative photoconductance (NPC), which opens up possibilities for high-speed computation with a faster on-off switching process, multi-bit memory, neuromorphic computing, and versatile sensor modes for temperature, electrostatic field and light wavelength within a single device configuration[16–19]. The co-integration of PPC and NPC phenomena can facilitate the implementation of logic gates and Boolean operations.[20] The observation of the NPC effect is mainly attributed to the mobility degradation originating from a range of factors.[17,21] Efforts are engaged to achieve electrical control over NPC-PPC through electrostatic doping and band tunneling within van der Waal heterostructure configurations.[19,22]

However, the exact microscopic origin of the NPC-PPC characteristics in 2D transistors has remained elusive. In this regard, the trion, with its higher effective mass, has been predicted as one of the major optical ensembles that can slow down the intrinsic conductivity of the material and, therefore, result in NPC in the terahertz conductivity.[23] Yet, the role of trions in tunable photoconduction remains to be established for designing multimodal optoelectronic devices.

In that context, 2D semiconductor lateral heterostructures (LHS) featuring spatially segregated n- and p-type domains[24–27] could provide a unique platform to investigate the accurate nature of these optical carriers, excitons and trions. The interfacial carrier lifetimes[28] and strong in-built asymmetry in the excitonic energy landscape across the one-dimensional heterointerface[2,29] within LHS are suitable for electrical modulation and unidirectional exciton transport.[28,30,31] These unique attributes make LHS-based excitonic devices intriguing for exploring their photoconductivity behavior under external bias and tunable light exposure; however, they have not been explored so far. We decipher the role of excitons and trions in the origin of negative photoconductance in 2D phototransistors. The ability to tune the photoconductance from negative to positive is demonstrated in 2D bilayer semiconductor n-p-n junctions. To probe the type of excitons and their complexes, the pump excitation wavelength ($\lambda_p$) was varied between 690 nm to 850 nm. The contribution from excitonic carriers is explained through a $\lambda_p$ and gate bias ($V_g$) dependent PPC to NPC transition across the lateral semiconductor interfaces. Moreover, a microscopic model has been developed to explain the experimental result for an in-depth analysis of the tunable photoinduced characteristics. Our findings facilitate designing novel photodetectors and light-driven memory and logic devices based on excitons for room temperature operation.

**Result and discussion**

2L LHS can provide an optimal balance between optical response and electrical carrier density, stability, minimal substrate effects, and low Schottky barrier height compared to a monolayer.[32] In this context, a multi-junction bilayer (2L) $MoSe_2$-$WSe_2$-$MoSe_2$-$WSe_2$ LHSs device (D-I) was fabricated with distinct 2D domains for illumination at $WSe_2$. The optical micrograph of the 3-junction LHS reveals the sequential in-plane deposition of individual TMDs by changing the carrier gases in situ via a robust CVD process, as per our earlier report (Fig. 1a).[24,25] Room temperature Raman spectral intensity line map across the lateral junctions highlights the sharp interface with distinct 2L $MoSe_2$ ($A_{1g}$ 240.2 cm$^{-1}$) and 2L $WSe_2$ ($A_{1g}$ 250 cm$^{-1}$) domains (Fig. 1b and S1a).[24,25] To confirm the lateral integration at the atomic level, we have carried out high-angle annular dark-field scanning transmission electron microscopy (HAADF-STEM) Z-contrast imagining that shows the presence of a distinct crystalline interface between spatially separated MoSe$_2$ and WSe$_2$ domains (Fig. 1c). The MoSe$_2$ domain appears darker compared to the WSe$_2$ domain due to their Z-number difference. The atomic arrangements indicate the bilayers are of the 2H$_c$ phase. The STEM and selected area diffraction images revealed that the as-synthesized 2D LHS are of high crystalline quality without any significant defects or strain across interfaces. The room temperature PL spectra from individual LHS domains show the emission peaks at 789 nm and 829 nm, which correspond to 2L-WSe$_2$ and 2L-MoSe$_2$ domains, respectively (Fig. 1d). We deliberately fabricated a device with a channel material of a narrow WSe$_2$ strip (~ 2 μm) between two MoSe$_2$ regions to investigate the transport of the exciton complexes from the WSe$_2$ to the MoSe$_2$ domain through the lateral interfaces. A thick hBN layer (~ 40 nm) was partially transferred onto the synthesized flake to access individual 2D TMD domains for electrical characterization (Fig. 1g). An external field bias was given to the MoSe$_2$ domains to directly observe critical interfacial phenomena, such as the formation, propagation, and diffusion of excitons. A proposed energy band diagram (Fig. S1c) of the n-p-n junction based on the barrier potential shows two depletion regions with opposite electric field directions, adjusted by the applied drain bias ($V_d$).

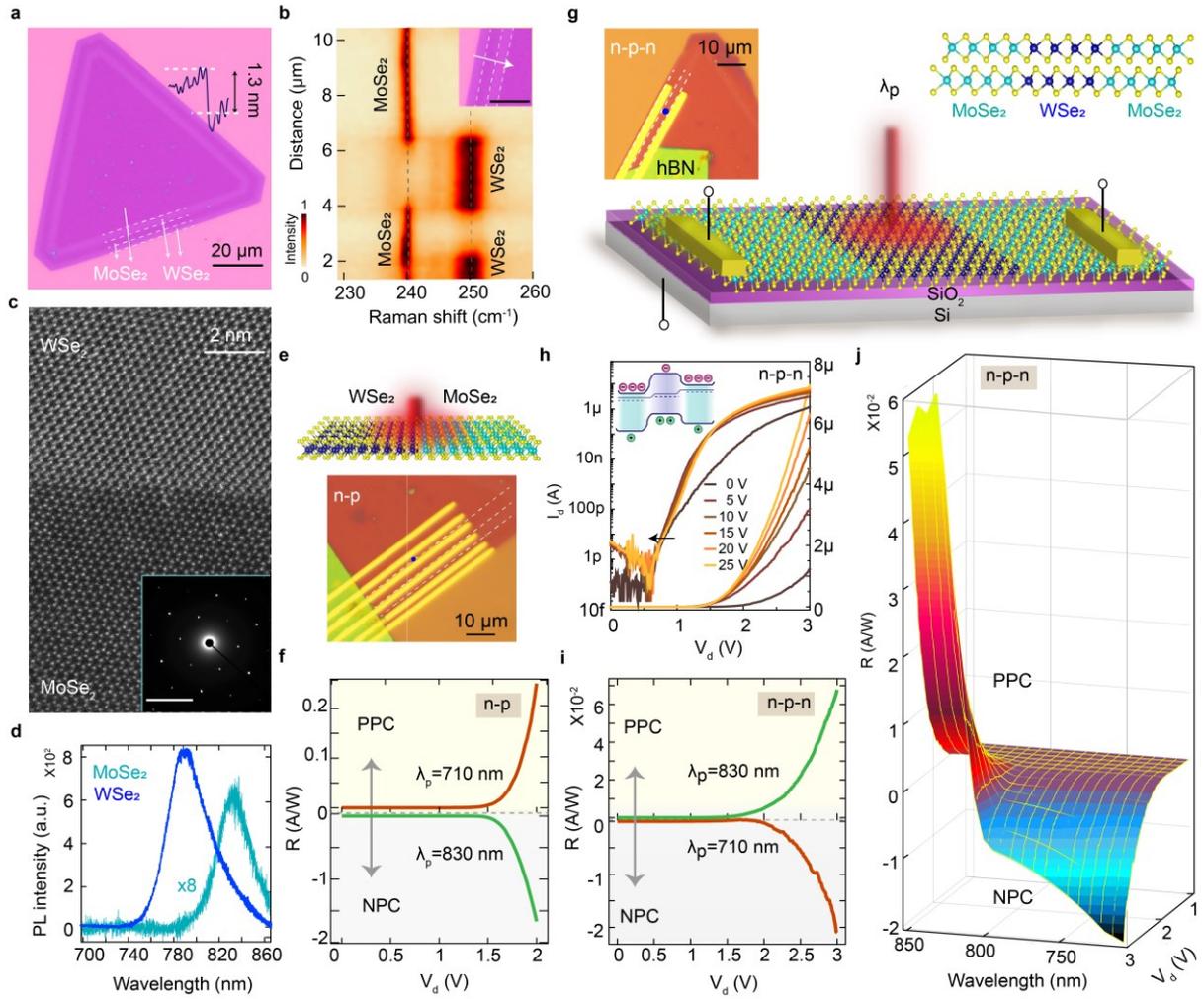

*Figure 1. Lateral 2D multi-junction heterostructures (LHS) device based on bilayer (2L) MoSe$_2$ and WSe$_2$. a. An optical image of a direct CVD-grown 2L three junction MoSe$_2$-WSe$_2$ LHS, where the colour contrast indicates the different domains. Inset, the AFM height profile shows the thickness of the 2L LHS. b. Raman spectral line scan across the interfaces within LHS showing the distinct crystalline domain and junction features extracted along the white arrow in the inset; The scale bar is 10 µm. c. Atomic-resolution HAADF-STEM image of a sharp 2L lateral MoSe$_2$-WSe$_2$ interface and the corresponding SAED pattern in the inset indicates the high crystallinity of as-grown LHS (scale bar is 5/nm). d. PL spectra of MoSe$_2$ and WSe$_2$ at room temperature showing the excitonic resonance wavelength at 829 and 789 nm, respectively. e. A schematic and the optical image of a one junction 2L n-p device and f. the corresponding photoresponse characteristics at the two different excitation wavelengths of 830 and 710 nm showing distinct negative and positive photoresponsivity (R). g. A schematic representation of the 2D lateral n-p-n device used for the study under variable pump-wavelengths ($\lambda_p$). The corresponding optical micrograph of the LHS device (left) and the ball model (right) shows the cross-sectional view of the 2L junctions. The laser beam (marked by a blue dot) only excites the WSe$_2$ region, while both the MoSe$_2$ domains are connected to the source and drain bias. h. The n-p-n output device characteristics at positive gate voltage ($V_g$) under dark conditions. Inset shows the corresponding band profile schematic indicating a shift in Fermi energy level ($E_F$) at high $V_g$ (25V). i. Photoresponse characteristics at the two different excitation wavelengths of 830 and 710 nm in n-p-n configurations show reverse photoresponse phenomena compared to the n-p device at $V_g$=0V. j. Variation of R with the tunable incident wavelength and applied source-to-drain bias ($V_d$) for the n-p-n device.*

Notably, the as-synthesized LHS with WSe$_2$ and MoSe$_2$ domains possess intrinsic hole and electron doped in nature, respectively, confirmed by their transfer characteristics (Fig. S2). In contrast to WSe$_2$, MoSe$_2$ exhibits relatively low gate modulation, indicating higher doping density. This results in a narrow depletion region at the MoSe$_2$ side, as the depletion width is inversely proportional to the doping density.[33] The output I-V characteristic in the dark depends on $V_g$ applied to the bottom silicon substrate is shown in Figure 1h and Figure S3a,b. A good gate-tuneability and low gate current (<30 pA) confirm the negligible substrate effect throughout the experiment (Fig. S3b inset). The ideality factor ($\eta$) 15.7 is determined from the I-V curve, fitted to the Shockley diode equation (Fig. S3c). Negative $V_g$ can increase the hole concentration in the WSe$_2$ domain while positive $V_g$ increases the electron density in the MoSe$_2$ domain. However, the Fermi level shifts in response to the external field depends on the doping densities.

The optoelectronic properties of the device were investigated at ambient conditions using a focused laser beam passed through an objective lens of 50X magnification (N.A 0.5), yielding a spot diameter of ~1.4 μm. A tunable laser is used for excitation in the WSe$_2$ domain of 2 μm width in the n-p-n configuration with the pump wavelength ($\lambda_p$) ranging from 690 nm to 850 nm. Upon irradiation, the absorption of a photon by the semiconductor triggers the generation of electron-hole pairs, wherein the mobile charge carriers subsequently contribute to the electric current. From the junction characteristic, photoresponsivity ($R$) is calculated to quantify the device's photoresponse. For the extraction of $R$ we used $R = I_{ph}/pA$; where $p$ denotes the power density of incident photons, and $A$ is the device active area. The maximum photocurrent ($I_{ph} = I_l - I_d$; $I_l$ is for illumination and $I_d$ is for dark) 12 µA (at 710 nm) and 28 µA (at 830 nm) are achieved from the n-p-n junction with a power density of 12.9 W/cm² and 6 W/cm², respectively, with a high signal-to-noise ratio of up to $10^6$. Here, the significance of the n-p-n device geometry lies in its ability to completely reverse the photoresponse characteristics compared to a single-junction (n-p) bilayer lateral phototransistor (Figs. 1e,f and i). In the n-p junction, high incident energy (710 nm) increases current due to the conventional increase of photogenerated carriers, while lower energy excitation (830 nm) reduces photocurrent due to the synergistic influence of two distinct types of trap carriers in WSe$_2$ and MoSe$_2$ (detailed in subsequent sections). Notably, in the n-p-n system, PPC emerges at 830 nm, while NPC is observed under 710 nm illumination. This contrast reveals distinct carrier dynamics within the n-p-n configuration, allowing selective excitation of energy states under the tunable incident wavelength (Fig. 1j).

Furthermore, a comprehensive study focusing on $\lambda_p$-dependent photocurrent measurement was conducted in the n-p-n in-plane phototransistor for a more in-depth understanding involving $V_g$ (Fig. 2). Photocurrent $I_{ph}$ and $R$ are plotted with varying $V_d$ for various $V_g$ corresponding to four different wavelengths lying between 710 to 830 nm (Fig. 2a-h). The observed photocurrent due to 830 nm laser excitation exceeds the dark signifying PPC, which increases by an increase in the applied $V_d$ for all the positive $V_g$ from 0 to 25 V. On the contrary, we observed a prominent tuning of PPC to NPC effect for a range of $V_d$ and $V_g$ for 810 nm and 790 nm laser illuminations. Further $\lambda_p$ of 710 nm, the $I_{ph}$ completely shifts to the NPC domain, even with the higher $V_g$. However, there is no signature of NPC for negative $V_g$ at positive $V_d$ irrespective of the incident wavelength (Fig. 2i-l). As discussed below and confirmed by our microscopic study (cf. Fig. 5), the PPC and NPC stem respectively from detrapping of excitons and formation of less-mobile trions, which cause an increase and a decrease of free mobile charges, respectively. The $\lambda_p$-dependence NPC and PPC phenomena can be understood by investigating the precise origin and formation dynamics of exciton complexes in the MoSe$_2$ and WSe$_2$ domains. The PL spectra reveal the existence of exciton and trion in MoSe$_2$ (Fig. S4a,b); however, resolving them is challenging at room temperature due to the thermal-induced broadening. PL spectra were recorded at 4 K from three different regions of the active channel materials to probe the exact nature of charge exciton further (Fig. 3a). The 2L MoSe$_2$ domain exhibits well-resolved emission peaks of neutral exciton ($X_0$) and trions ($X^-$) at 763 and 774 nm, respectively, with a separation of 11 nm (~23 meV). The broad emission peak at 893 nm can be attributed to indirect gap emission or mid-band defects.[34,35] The PL spectrum at 2L WSe$_2$ reveals a strong trap-related narrow emission band, ranging from 790 to 828 nm, while the exciton band is at 729 nm. The PL spectrum at the interface exhibits a combined signature from MoSe$_2$ and WSe$_2$ domains, revealing exciton states and defect or trap states within both materials. Power-dependent PL spectra confirm trion predominately appearing in the MoSe$_2$ domain and exhibiting conventional power-law variation characteristics (Fig. 3b,c).

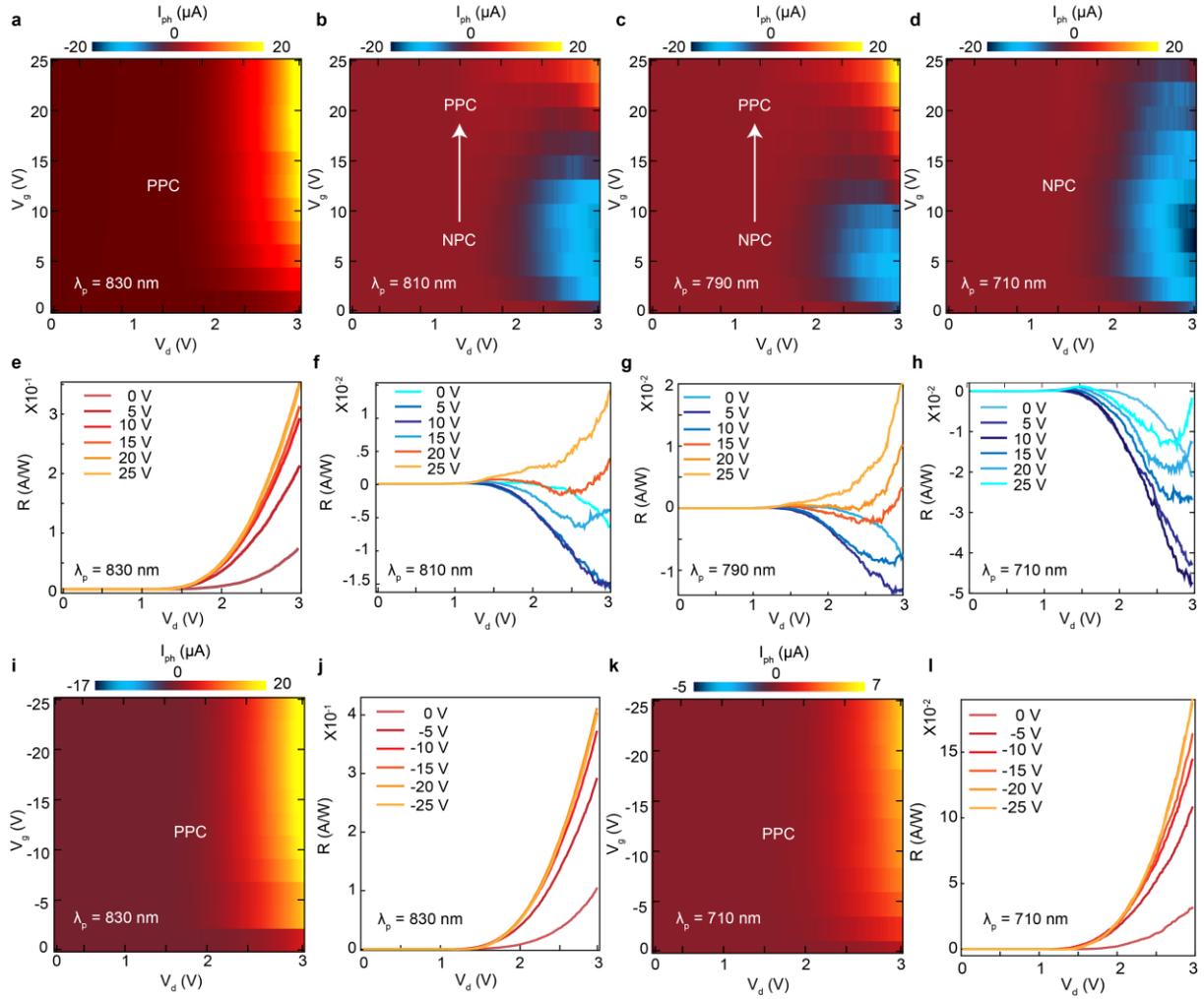

*Figure 2. Wavelength-dependent NPC-PPC transition. a-d.* 2D contour plots of the obtained photocurrent ($I_{ph}$) of the n-p-n device under illumination for the different wavelengths when $V_d$ and $V_g$ varies from 0 to 3 and 0 to 25 V, respectively. It shows the sequence of PPC to NPC transition that depends on both the $V_g$ and the excitation wavelengths. The color bars represent the extracted photocurrent range from the device. *e-h.* The corresponding photoresponsivity (R) with varying $V_d$ at different $V_g$. Another set of 2D contour plots and corresponding R-profile with varying $V_d$ at -$V_g$ varies from 0 to -25 V for *i, j.* 830 nm and *k, l.* 710 nm incident wavelength representing PPC phenomena.

Notably, complete NPC and PPC phenomena occur well below and above the resonant emission wavelength (790 nm) of the optical active WSe$_2$ domain at room temperature. Excitation wavelength larger than the exciton resonance (bandgap wavelength) cannot form an electron-hole bound state (free charge carrier) in 2D TMDs. Therefore, the observed PPC characteristic for below resonant excitation ($\lambda_p$>789 nm) can be ascribed to the presence of low-lying trap states ($E_T$) below the conduction band energy in WSe$_2$, as it is the only photo-excited domain. Under the non-equilibrium condition, the minority carriers can be trapped to $E_T$.[36] Below resonant excitation, these trapped charge carriers can jump to the conduction band, increasing photoconductance. Hence, a positive photoresponse is expected, as indicated in the diagram (Fig. 3d). In the WSe$_2$ domain, with increasing temperature, trap-assisted emission bands are redshifted and extend up to 850 nm at room temperature (Fig. S4c). Consequently, even with optical excitation up to 850 nm, we anticipate PPC, as confirmed by photocurrent measurements (Fig. S5). Thus, NPC-PPC optoelectrical phenomena can determine the optimum energy range of the effective trap states in WSe$_2$ and trion in MoSe$_2$, even at room temperature.

The photon energy above the band edge excites electrons from both trap levels and the valence band to the conduction band. In this case, the resonant excitation of 790 nm is sufficient to generate photoexcited electron-hole bound pairs. Further lowering $\lambda_p$ induces the formation of the high density of exciton in WSe$_2$ in addition to free electrons and holes in the conduction and valence band, respectively. The photogenerated exciton in WSe$_2$ can transmit towards the MoSe$_2$-WSe$_2$ interface owing to the narrow channel width of the WSe$_2$ domain (2 µm).[28] Subsequently, it can diffuse to MoSe$_2$ across the interface due to the potential anisotropy.[28,31] Simultaneously, applying positive $V_g$ increases the free electrons density as a result of electrostatic doping on n-type MoSe$_2$. The diffused excitons from the WSe$_2$ domain can interact and bind with electrons near the interface, resulting in the formation of charged excitons or trions (Fig. 3f). The trion possesses high dissociation energy and high effective mass, which makes them less mobile, leading to reduced photocurrent and hence the transition to the NPC regime.[23] The interplay between the density of trap states and the formation of trion results in a transitional region from PPC to NPC within the range from 790 to 810 nm (Fig. 3e).

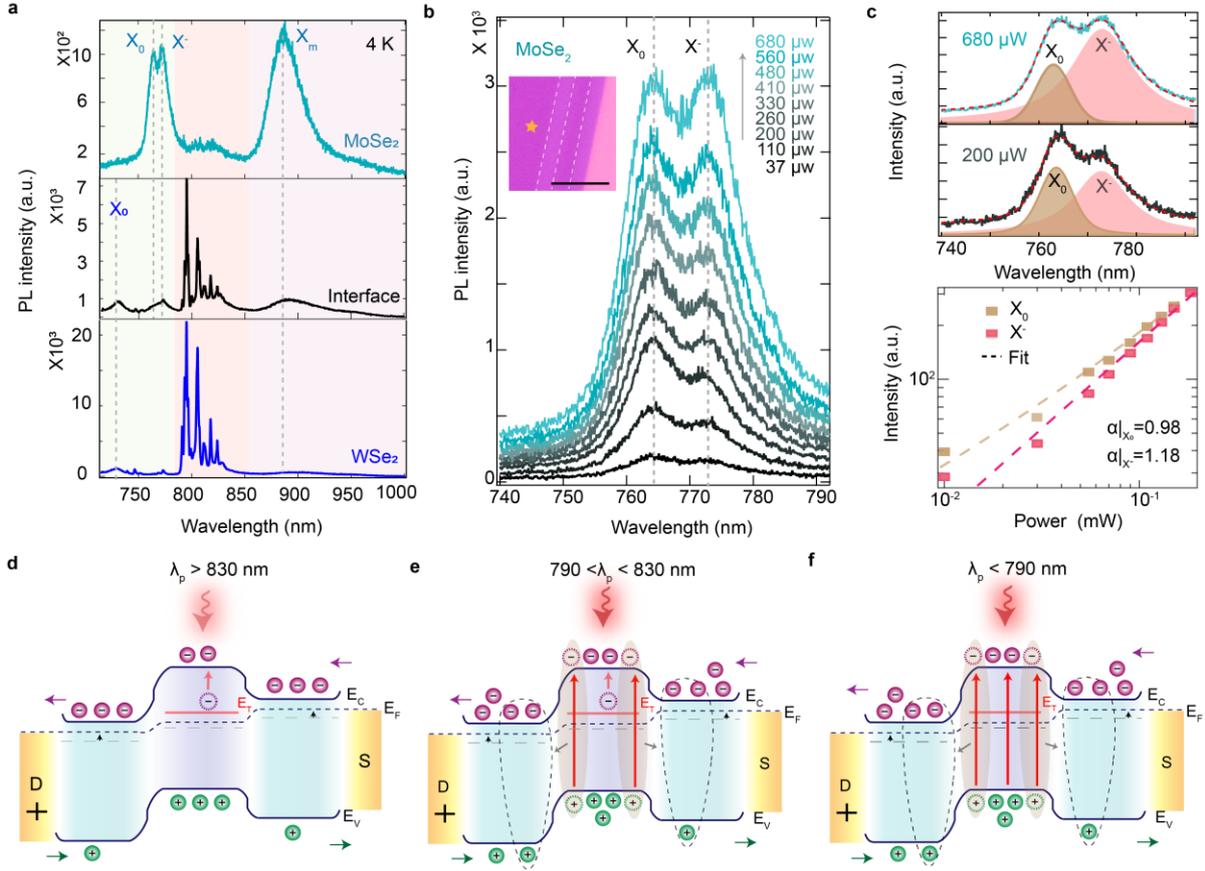

*Figure 3. Low-temperature optical characteristic of LHS and charge transfer mechanism. a. Low-temperature (4 K) PL spectrum of MoSe$_2$, interface, WSe$_2$; MoSe$_2$ domain possesses a clear trion and exciton emission with a broad defect peak at 893 nm (FWHM: 44 nm), whereas WSe$_2$ domain shows strong, sharp emission from defect or trap states; The color shades indicate the different wavelength regions associated with neutral and charged excitons as well as defect related emissions for both the material domains. b,c. The power dependence PL spectra at MoSe$_2$ (marked in the inset) show the increase in the trion peak with power while the peak position remains invariant. The PL peak fit indicates the change in the trion and exciton intensity ratio. Scale bat in b inset is 10 µm. d-f. Proposed charge transfer mechanism with energy band diagram for three different laser excitation regimes (at $V_g$ = 25 V) at positive bias mode. d. For $\lambda_P$ > 830 nm, the trapped carrier participates in the photocurrent. e. For 790 < $\lambda_P$ < 830 nm, both the trap carriers and photogenerated exciton in WSe$_2$ participate in the PPC to NPC transition. f. For $\lambda_P$ < 790, photogenerated excitons diffuse and form trion to the adjacent MoSe$_2$ domain by capturing free electrons, resulting in NPC.*

The PPC characteristics over an extended range of applied negative $V_g$ are related to the reduction of the free carrier in MoSe$_2$ (Figs. 2i-l and Fig. S6). Particularly when the device is in an on-state with $V_g$<0, the holes accumulate in the WSe$_2$ side, and electrons deplete in MoSe$_2$, resulting in a reduced probability of trion formation in the MoSe$_2$ domain. This indicates the contribution of trion on photocurrent is predominantly determined by the majority carrier in MoSe$_2$. Moreover, negative $V_g$ can effectively enhance the concentration of holes in the valence band of WSe$_2$, resulting in a high PPC throughout all the excitation $\lambda_p$. However, photoexcitation injects additional charge carriers in WSe$_2$ for $\lambda_p$ lower than the resonance excitation, increasing the current through conventional photon absorption and electron−hole generation. Trions associated with NPC can be assigned as negative since they exhibit a maximum response only at positive $V_g$. While interface excitons have recently been observed,[26,37] the formation of interface trions at lateral junctions could also lead to the NPC effect. Various mechanisms exist for interlayer trion formation: 1) a bound state with an electron residing in MoSe$_2$ and one photogenerated exciton in WSe$_2$ or 2) a photogenerated hole in WSe$_2$ can be transferred to MoSe$_2$ and thus interface trion can form with two electrons in MoSe$_2$ and one hole from WSe$_2$. Yet, our recent findings confirm that the concentration of trions lies within the MoSe$_2$ domain near the interface rather than forming interfacial trions, which can mitigate these two possibilities.[38] On another note, the majority carrier electrons from MoSe$_2$ can diffuse under non-equilibrium conditions and form trions by capturing a photogenerated exciton within the WSe$_2$ domain. However, this is less likely to be possible in 2L WSe$_2$ due to the presence of trap states, which can neutralize the free charge carriers through recombination instead of trion formation.

The wavelength-dependent responsivity is plotted over varying $V_g$ at a particular $V_d$, where the R follows a PPC behaviour over a range of negative $V_g$ and switching from NPC to PPC at positive gate bias (Figs. 4a and S7). This reinforces the junction-dependent optical feature, as the NPC and PPC phenomena in single junction devices are consistent under both forward and reverse bias conditions, as well as with positive and negative $V_g$ (Fig. S8). Thus, while the excitation wavelength is the only tool for switching between NPC and PPC in the single n-p junction LHS, the n-p-n configuration offers an additional degree of freedom through both drain and gate bias.

The detectivity is a crucial performance parameter for a photodetector. Fig. 4b shows the specific detectivity ($D^*$) limited by short noise for different incident wavelengths with varying $V_g$ at 3 V of $V_d$. The $D^*$ is defined by $R(A/2qI_d)^{1/2}$, where q is the elemental charge. The two-junction device demonstrates a maximum $D^*$ value of $1.92 \times 10^{10}$ Jones, comparable with the reported values for exfoliated and synthesized TMDs within monolayer to multilayers.[39] Furthermore, the charge carrier transfer characteristics were successfully probed in the n-p-n junction by controlling the photon injection density. Two distinct photoresponse regions are observed for excitations below and above resonance wavelength (Fig. 4c). The device's responsivity to the 710 nm laser increases almost linearly with $P$. In contrast, the response to 840 nm excitation gradually saturates at high power. It supports the proposed trion and defect-induced photoconductance phenomena as the trion density increases and defect states become saturated with increasing power.

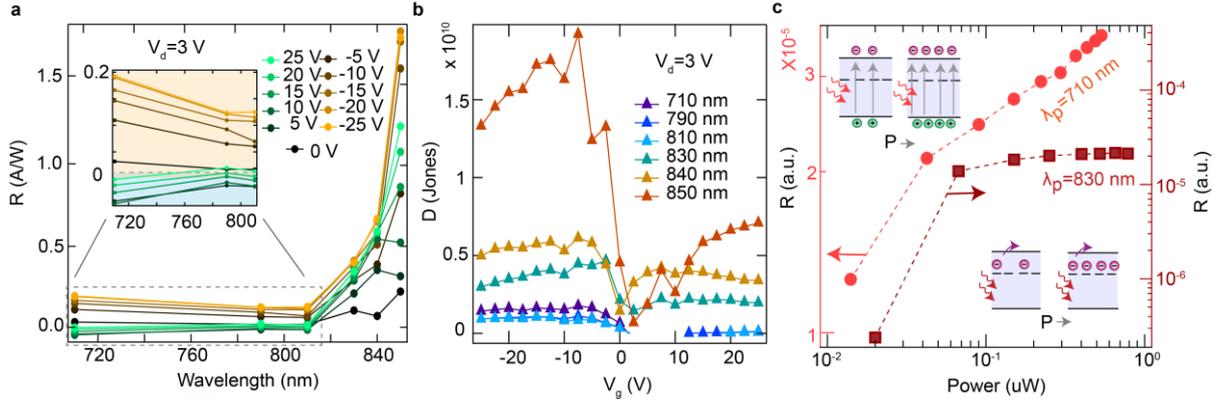

*Figure 4. Wavelength-dependent optoelectrical characteristics of 2D LHS FET. **a.** The wavelength-dependent responsivity of the n-p-n junction is shown at $V_d=3$ V. R for $+V_g$ shows NPC, and this follows as PPC for $-V_g$. An enlarged image of the photoresponsivity profile is shown in the inset to distinguish the NPC-PPC clearly. **b.** Variation of specific detectivity with $V_g$ at $V_d=3$ V under different wavelength irradiation. **c.** Power dependence photocurrent in a log-scale (in an arbitrary unit) at two different excitation wavelengths at $V_g = 0$ V. The 710 nm excitation shows almost linear variation of photoresponse with power, while at 830 nm, it saturates with increasing the laser power. The possible carrier generation mechanisms are shown in the inset for the same two wavelengths with increasing power.*

Upon illumination, electrons and holes are generated more efficiently in the large bandgap material WSe$_2$ in this device geometry. Therefore, the crossover wavelength for the NPC to PPC transition may vary slightly between 780−800 nm across devices, depending on the exciton resonance of CVD-grown WSe$_2$. The reproducibility of the findings was confirmed with three additional devices with n-p-n geometry (Fig. S9). To rule out the Schottky barrier effect impacting gate-dependent tunable photoconduction due to metal contacts on the TMD domains, we fabricated two devices with different metals, Ti/Au (5/70) contacts (D-II and D-III), while D-I and D-IV were fabricated with Ti/Pt/Au (5/30/65). We observed consistent photoconductive behavior in all the devices (D-II, III and IV) across $\lambda_p$ with tuneable $V_g$, demonstrating the similar NPC and PPC phenomena (Fig. S9). Also, the calculated photogain [$G = Rhc/(\lambda n e)$, where $n$ is the external quantum efficiency] corresponding to different $V_g$ shows the wavelength dependence nature for the n-p-n devices (Fig. S10). Thus, we establish a comprehensive investigation into a robust tuneable photocurrent generation through lateral interfaces involving the roles of exciton, trion, and trap states.

**Microscopic modeling**

To understand the above experimental result, we microscopically investigate the interplay of gate-controlled doping and optical excitation. Resonant laser pulses excite excitons, which then bind with free electrons or holes for the formation of trions. This results in a trion-induced decrease of the current via a reduction of free charges, Fig. 5a. We microscopically describe the relative density of charges $n_c$, excitons $n_X$ and trions $n_T$ by evaluating the Saha equation[40]

$$\frac{n_c n_X}{n_T} = \frac{m_c m_X}{m_T} \frac{k_B T}{2\pi \hbar^2} e^{-\frac{E_{b,T}}{k_B T}} \qquad (1)$$

where $m_c$, $m_X$ and $m_T$ are the mass of holes, excitons, and trions, respectively. Furthermore,

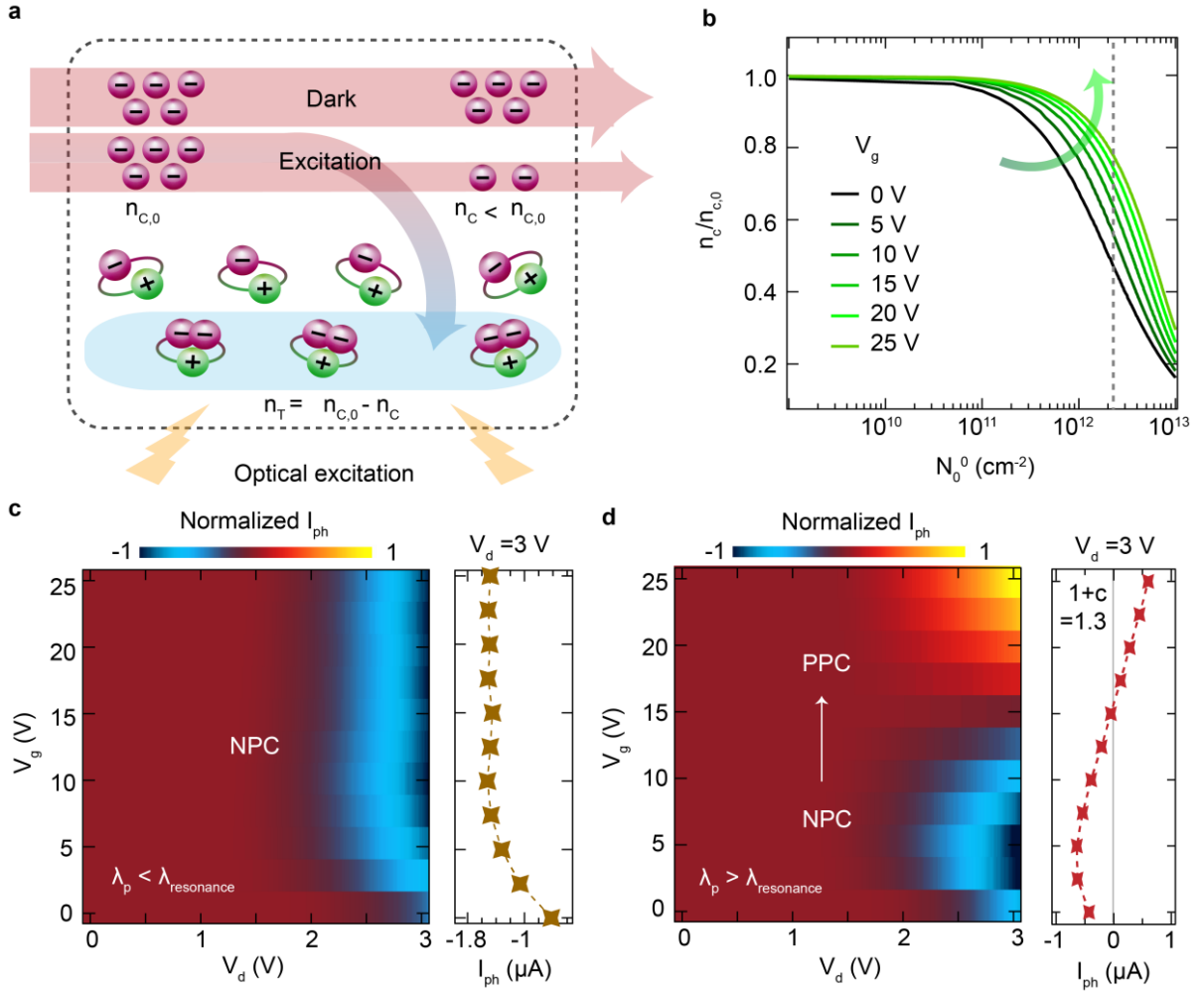

*Figure 5. a. Theoretical validation of NPC-PPC effects. **a.** A schematic depicting the optical excitation-induced carrier density and formation of trions utilized for theoretical simulations, resulting in a decrement in the overall photocurrent. **b.** The free carrier and gate-dependent doping density ratio profile with the increasing total carrier density $N_0^0$ inside the system at different $V_g$ based on the theoretical model. **c.** The 2D contour plot illustrates the variation in simulated photogenerated current with $V_d$ and $V_g$ showing NPC at a power density of 12.9 W/cm$^2$ ($\lambda_p$=710 nm); A line profile of $I_{ph}$ with $V_g$ for a fixed $V_d$=3 V (right) demonstrating saturation behavior at 7.5 V onwards. **d.** A 2D contour plot for photogenerated current shows NPC to PPC transition while $\lambda_p$ is set above the resonance excitation. A line profile of $I_{ph}$ at 3 V of $V_d$ (right) considering 30% of de-trapping (1 + c = 1.3) demonstrates the NPC to PPC transition over the $V_g$.*

$T$ is the temperature and $E_{b,T}$ is the binding energy of trion, cf. Methods. Equation (1) is coupled with,

$$n_C + n_T = n_{c,0}(V_g), \text{ and} \qquad (2a)$$

$$n_X + n_T = N_0^\circ, \qquad (2b)$$

which impose the conservation of charges and optically-excited particles, respectively. Here $n_{c,0}$ and $N_0^\circ$ are the density of free and excited particles, respectively, which can be experimentally controlled via gate and laser power, cf. Methods. Equation (2a) shows how trions reduce the density of free-charges from $n_{c,0}(V_g)$ to $n_C < n_{c,0}(V_g)$, resulting in a decrease of the current after optical excitation, as sketched in Fig. 5a. Such a reduction $\frac{n_c}{n_{c,0}} < 1$ depends on both gate potential $V_g$ and excitation density $N_0^\circ$, as shown in Fig. 5b. While the free charges are unaffected by weak excitation, $\frac{n_c}{n_{c,0}} \approx 1$ for $N_0^\circ \lesssim 10^{11} \text{cm}^{-2}$, for larger excitation densities, the ratio $\frac{n_c}{n_{c,0}}$ drops, in particular for smaller gates/doping $n_{c,0}(V_g)$, resulting in a higher relative amount of charges bound into trions (black vs green). Notably, at the excitation density $N_0^\circ = 2.2 \times 10^{12} \text{cm}^{-2}$ corresponding to the experiments (vertical dashed lines), all the values are well below 1.

After having addressed how gate and laser power affect the relative reduction of free charges $\frac{n_c}{n_{c,0}}$, we can evaluate the current after illumination, $I_l = \frac{n_c}{n_{c,0}} I_d$ by taking the dark current $I_d$ extracted from the experiment (Fig. S3a). This results in the photocurrent

$$I_{ph} = I_l - I_d = I_d \frac{n_c - n_{c,0}}{n_{c,0}} \equiv I_d \left( \frac{n_c}{n_{c,0}} - 1 \right),$$

whose values are shown in Fig. 5c for the case of resonant excitation. We predict the appearance of NPC in good agreement with experiments at 710 nm, Fig. 2d. Despite the increase of $I_d$ with $V_g$, cf. Fig. S2a, surprisingly, we theoretically predict a saturation of NPC for gates increasing beyond 5-10 V, which agrees with the measured behaviour, cf. Fig. 2d. To better show this, in the side panel, we show $I_{ph}$ as a function of $V_g$ at a fixed bias of $V_d = 3$ V. Such a saturation for increasing gates of $I_{ph}$ stems from the increase of the ratio $\frac{n_c}{n_{c,0}}$ (arrow and black to green lines in Fig. 5b), resulting in a decrease of $\frac{n_c}{n_{c,0}} - 1 = \frac{n_c - n_{c,0}}{n_{c,0}}$ which compensates for the increase of $I_d$. In Fig. 5d, we consider optical excitations energetically below the resonance ($\lambda_p > 789$ nm).

Here, the trion formation of Eqs. (1-2) is in competition with an increase of doping from $n_{c,0}(V_g)$ to $n'_{c,0}(V_g) = n_{c,0}(V_g)(1+c)$ induced by exciton de-trapping, where $c$ indicates the relative increase of free charges. Including this competition between de-trapping and trion formation, we obtain an NPC to PPC transition with increasing $V_g$, Fig. 5d, in agreement with the experiments (Fig. 2c).

**Summary**

We demonstrated optical and electrical pathways to precisely control the photoconductivity behavior in CVD-grown bilayer n-p-n lateral heterostructure (LHS) FET devices. The unique in-plane geometry in 2D LHS not only provides built-in potential asymmetry for spontaneous exciton diffusion but also offers a new degree of freedom, enabling domain-specific carrier generation and propagation across the interface. We demonstrated the switching of positive to negative photoconduction (NPC to PPC) effect for the first time in 2D LHS through the formation and electrical manipulation of trions. The combined influence of trions in $MoSe_2$ and trap states in $WSe_2$ significantly alters the transition between NPC and PPC over a tunable pump wavelength range of 690 nm to 850 nm. In addition, a microscopic model incorporating exciton, trion and doping density shows the free carrier reduction in charge transport upon resonant illumination due to trion formation, whose competition with de-trapping results in a gate-induced transition from NPC to PPC for excitation energies below resonance. We achieved passive realization of exciton, trion and trap carrier effects on the tunable photoconduction of a 2D lateral heterostructure FET device, which is otherwise challenging to resolve with the optical probing at room temperature. These findings on precise control of optical carriers and optoelectrical features across one-dimensional lateral 2D semiconductor heterointerfaces can pave the way for designing atomically thin exciton-based transistors, logic gates, memory and sensors. Further exploration into exciton and trion dynamics, including their formation and dissociation energy, localization, recombination, and annihilation, can significantly enhance the understanding of future advancements in quantum information processing, quantum communication, and energy conservation.

**Materials and methods**

**Synthesis of 2L $MoSe_2$-$WSe_2$ LHS**

The three-junction 2L $MoSe_2$-$WSe_2$ LHSs were synthesized using a one-pot water-assisted CVD method at 1060°C with $MoSe_2$ and $WSe_2$ bulk precursors.[24,25] By changing in situ carrier gases, we achieved selective growth of different TMD layers while maintaining substrate temperatures at 800 °C. Initially, wet N$_2$ gas with 200 s.c.c.m. oxidized solid precursors, promoting Mo-based precursor evaporation to form MoSe$_2$. Switching to a reducing gas (H$_2$+Ar) allowed W-related sub-oxides to evaporate, facilitating the formation of WSe$_2$ at the same temperature. By alternating the gas flow cyclically, continuous multi-junction LHSs have been fabricated while maintaining independent control over the lateral dimensions, shape, and size of individual domains.

**Device fabrication**

To fabricate the FET geometry with the LHS synthesized on 285 nm thick SiO$_2$/n-doped Si, a thick hBN flake was selectively transferred using a scotch tape-based dry exfoliation technique and partially overlaying the as-grown LHS flakes. The hBN flake acts as an insulator beneath the metal electrodes connecting to the individual center and second MoSe$_2$ domain. Next, source and drain contacts were fabricated using electron beam lithography (Raith 150-Two) on MoSe$_2$ and WSe$_2$ individually. Finally, source/drain metallization- either 5 nm Ti/ 30 nm Pt/ 65 nm Au or 5 nm Ti/ 70 nm Au via sputtering process followed by lift-off was used to fabricate FET devices.

**Experimental setup and procedure for TEM, optical and optoelectrical measurement**

For the HAADF-STEM sample preparation, we have carried out standard PMMA-assisted wet transfer of the as-grown TMD flakes to the TEM grid. HAADF-STEM imaging was performed in an aberration-corrected JEOL JEM-ARM300F2 microscope with a cold-field emission gun at 300 kV and JEOL HAADF detector using the probe size 8c and scan speed 20 μs per pixel.

The PL measurements were conducted in the temperature range of 4 K to RT using a closed-cycle optical microscopy cryostat (Montana Instruments). PL signals were collected using Princeton Instruments spectrometer (SP2750) and a liquid nitrogen-cooled charge-coupled detector (PyLoN:400BR-eXcelon) dispersed by a diffraction grating with 1200 grooves per mm (spectral resolution of 0.02 meV). A CW laser with a pump wavelength of 660 nm was focused to a spot size of ~1 μm by a 60X objective lens (N.A. = 0.7).

For the optoelectrical measurements, the contact electrodes of the device were wire bonded using gold wire to the PCB metal pads. The measurements were carried out on an Olympus microscope (Model BX-63) using a semiconductor device analyzer (Keysight B1500A) and a tunable laser (NKT Photonics SuperK SELECT) in the wavelength range of 690 to 900 nm. The laser was incident on the device through a 50X objective lens (N.A.-0.5).

**Microscopic methods**

Our theoretical predictions are obtained via the microscopic solution of the Saha equation, where the effective masses ($m_c = 0.27 m_0$, $m_X = 0.63 m_0$ and $m_T = 0.9\ m_0$ with $m_0$ being the free electron mass) are taken from DFT calculations.[41] Furthermore, we assume room temperature and $E_{b,T}$=20 meV as the trion binding energy - in agreement with our measurement in bilayer MoSe$_2$ (Fig. 3d) and reported values in monolayers.[42,43] We neglect the direct excitation of free electron-hole pairs, as this is negligible given the high binding energies and presence of doping.

The Saha equation (Eq. (1)) has to be coupled with Eqs. (2a-b), which impose the conservation of charges and excited particles, respectively. The doping $n_{c,0}(V_g)$ in Eq. (2a) depends on gate voltage $V_g$ via $n_{c,0}(V_g) \equiv n_{c,0} = n_c^0 + r_V V_g$, where $n_c^0$=3×10$^{11}$ cm$^{-2}$ and $r_V$=3×10$^{11}$ cm$^{-2}$/V are, respectively, the intrinsic doping and the rate of doping variation per applied Volt found in monolayer WSe$_2$.[44] In Eq. (2b), the equilibrium density of optically excited particles (excitons or trions) $N_0^\circ = x \frac{p}{\hbar \omega_l} \tau$ is given by the competition between excitation and decay and is found by applying the equilibrium condition $d_t N_0 \equiv d_t N_0^\circ = 0$ to the dynamics of the excited particles $d_t N_0 = x \frac{p}{\hbar \omega_l} - \frac{N_0}{\tau}$. Here, we find $N_0^\circ$=2.2 × 10$^{12}$ cm$^{-2}$ by using the power density $p$ =12.9 W/cm$^2$, measured at the incident photon energy of $\hbar \omega_l$ = 1.746 eV (710 nm). were $x \approx 0.1$ is the ratio of absorbed photons and $\tau$ = 450 ns is the typical decay times measured in the slow component of the bi-exponential decay in heterobilayers.[45–48]

**Acknowledgement**

PS acknowledges the Department of Science and Technology (DST) (Project Codes: DST/NM/TUE/QM-1/2019; DST/TDT/AMT/2021/003 (G)&(C)), India. BK acknowledges Pankaj Kumar Gound, IIT Bombay, for wire bonding. TEM work was performed at Sophisticated Analytical Technical Help Institute (SATHI), IIT Kharagpur, supported by DST, Govt of India. PS and SC acknowledge Mr. Arup Ghoshal for STEM measurement at SATHI. SL acknowledges funding from the DST, India (Project Code: DST/NM/TUE/QM-8/2019 (G)/2. We also acknowledge the Indian Institute of Technology Bombay Nanofabrication Facility (IITBNF) and Indian Nanoelectronics User Program (INUP) for the device fabrication and characterization. SD acknowledge SERB (CRG/2018/002845) and MoE (MoE/STARS- 1/647). SPD acknowledges the 2D TECH VINNOVA competence center at CHALMERS, Sweden (No. 2019-00068). EM and RR acknowledge financial support from the Deutsche Forschungsgemeinschaft (DFG) via SFB 1083 (project B9) and SPP 2244, and they thank R. Perea-Causin for helpful discussions.

**Additional information**

**Supplementary information:** The supplementary material available online

*Present Address of CS: Department of Electrical & Computer Engineering, University of Delaware, Newark, Delaware 19716, United States

30. Lamsaadi, H. *et al.* Kapitza-resistance-like exciton dynamics in atomically flat MoSe$_2$-WSe$_2$ lateral heterojunction. *Nat. Commun.* **14**, 5881 (2023).
31. Shimasaki, M. *et al.* Directional Exciton-Energy Transport in a Lateral Heteromonolayer of WSe$_2$ –MoSe$_2$. *ACS Nano* **16**, 8205–8212 (2022).
32. Fang, M. *et al.* Controlled Growth of Bilayer-MoS$_2$ Films and MoS$_2$-Based Field-Effect Transistor (FET) Performance Optimization. *Adv. Electron. Mater.* **4**, 1700524 (2018).
33. Choi, W. *et al.* Optoelectronics of Multijunction Heterostructures of Transition Metal Dichalcogenides. *Nano Lett.* **20**, 1934–1943 (2020).
34. Chen, K. *et al.* Experimental evidence of exciton capture by mid-gap defects in CVD grown monolayer MoSe2. *Npj 2D Mater. Appl.* **1**, 1–8 (2017).
35. Arora, A., Nogajewski, K., Molas, M., Koperski, M. & Potemski, M. Exciton band structure in layered MoSe2: from a monolayer to the bulk limit. *Nanoscale* **7**, 20769–20775 (2015).
36. Yin, L. *et al.* Robust trap effect in transition metal dichalcogenides for advanced multifunctional devices. *Nat. Commun.* **10**, 4133 (2019).
37. Yuan, L. *et al.* Strong Dipolar Repulsion of One-Dimensional Interfacial Excitons in Monolayer Lateral Heterojunctions. *ACS Nano* **17**, 15379–15387 (2023).
38. Kundu, B. *et al.* Electrically Controlled Interfacial Charge Transfer Induced Excitons in MoSe$_2$-WSe$_2$ Lateral Heterostructure. Preprint at http://arxiv.org/abs/2407.13724 (2024).
39. Shin, G. H., Park, C., Lee, K. J., Jin, H. J. & Choi, S.-Y. Ultrasensitive Phototransistor Based on WSe$_2$–MoS$_2$ van der Waals Heterojunction. *Nano Lett.* **20**, 5741–5748 (2020).
40. Quick, M. T., Ayari, S., Owschimikow, N., Jaziri, S. & Achtstein, A. W. Quantum Nature of THz Conductivity: Excitons, Charges, and Trions in 2D Semiconductor Nanoplatelets and Implications for THz Imaging and Solar Hydrogen Generation. *ACS Appl. Nano Mater.* **5**, 8306–8313 (2022).
41. Kormányos, A. *et al.* k · p theory for two-dimensional transition metal dichalcogenide semiconductors. *2D Mater.* **2**, 022001 (2015).
42. Courtade, E. *et al.* Charged excitons in monolayer WSe$_2$: Experiment and theory. *Phys. Rev. B* **96**, 085302 (2017).
43. Mak, K. F. *et al.* Tightly bound trions in monolayer MoS$_2$. *Nat. Mater.* **12**, 207–211 (2013).
44. Wagner, K. *et al.* Diffusion of Excitons in a Two-Dimensional Fermi Sea of Free Charges. *Nano Lett.* **23**, 4708–4715 (2023).
45. Kulig, M. *et al.* Exciton Diffusion and Halo Effects in Monolayer Semiconductors. *Phys. Rev. Lett.* **120**, 207401 (2018).
46. Varghese, S. *et al.* Fabrication and characterization of large-area suspended MoSe$_2$ crystals down to the monolayer. *J. Phys. Mater.* **4**, 046001 (2021).
47. Cai, C.-S. *et al.* Ultralow Auger-Assisted Interlayer Exciton Annihilation in WS$_2$/WSe$_2$ Moiré Heterobilayers. *Nano Lett.* **24**, 2773–2781 (2024).
48. Zhao, S. *et al.* Excitons in mesoscopically reconstructed moiré heterostructures. *Nat. Nanotechnol.* **18**, 572–579 (2023).
18

## Supplementary Note 1

### Band diagram

The junction potential is estimated as 60 meV via a Kelvin probe force microscopy (KPFM) measurement under dark conditions (Figure S1). The thermal equilibrium state of the MoSe$_2$-WSe$_2$-MoSe$_2$ band diagram is given in Figure 1a, considering type-II band alignment. $\Phi$ and $\chi$ denote the work function and electron affinity. $E_g$, $E_0$, $E_C$, $E_F$, and $E_V$ are the band gap energy, vacuum energy, minimum conduction band energy, Fermi energy, and maximum valance band energy levels, respectively. $\Phi$ value for the metal and $\chi$ values for WSe$_2$ and MoSe$_2$ are used as 5.3 eV, 3.91 eV, and 3.61 eV, respectively. The bandgap for MoSe$_2$ and WSe$_2$ is calculated as 1.62 and 1.68 eV, respectively. The lateral widths of MoSe$_2$ and WSe$_2$ are taken as 3 and 2 µm, respectively, where the metal contacts are placed on 2x2 µm of MoSe$_2$, as shown in Figure S4. The vertical width of 1.2 nm each is considered, and the relative permittivity is 7 and 7.2, respectively.

### Supplementary Figure

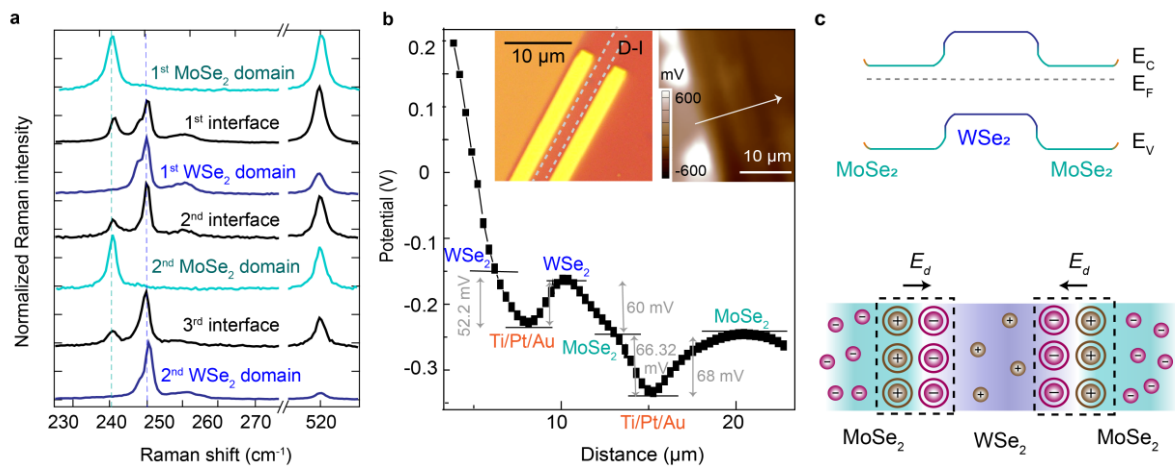

*Figure S1. **a.** Raman spectra of WSe$_2$, MoSe$_2$, and at the interface from the outer WSe$_2$ domain (2$^{nd}$ domain) to the centre. **b.** The KPFM surface potential profile across the various junctions indicated in the inset. **c.** A typical energy band diagram at equilibrium and a schematic of the channel material features two junctions representing p-type and n-type domains, along with mobile and immobile charges. The depletion regions, marked by dashed lines, exhibit opposite field directions.*

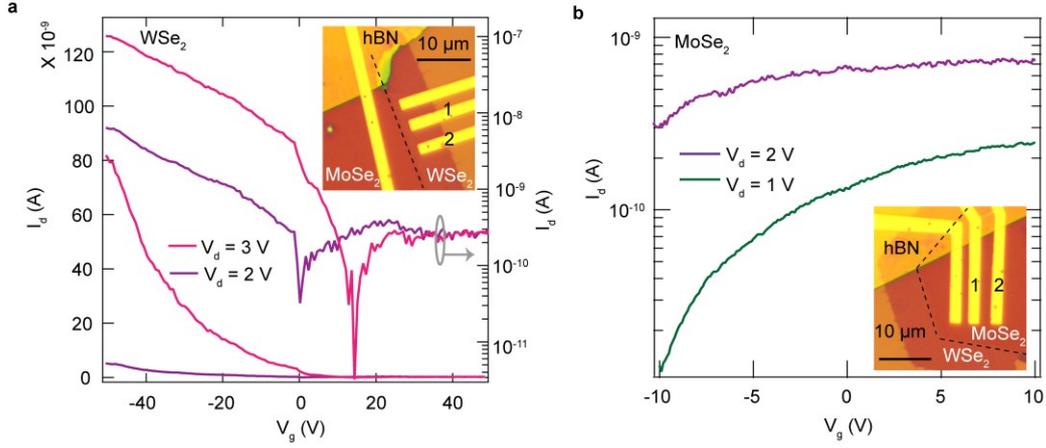

***Figure S2. a,b.*** *The transfer characteristics of devices where WSe$_2$ and MoSe$_2$ are the individual channel materials show p-type and heavily doped n-type characteristics.*

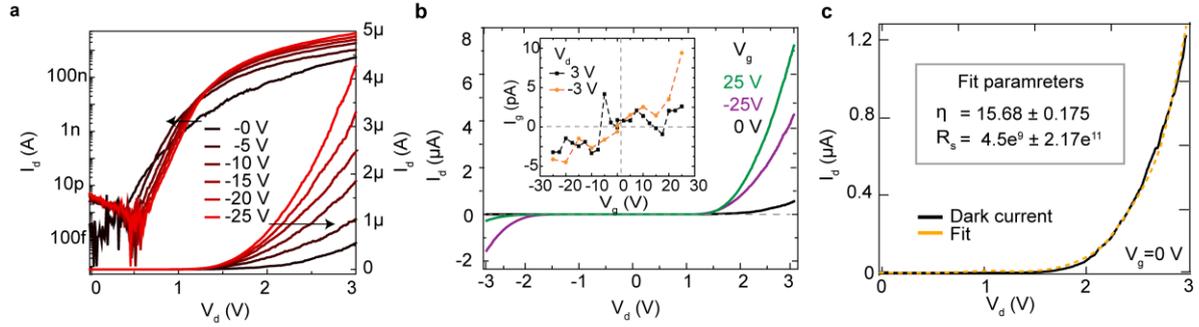

***Figure S3. a,b.*** *The output characteristics of 2-junction MoSe$_2$-WSe$_2$-MoSe$_2$ lateral heterostructure device at negative $V_g$ under dark conditions.* ***b.*** *I-V characteristics in dark conditions with the three different gate biases showing the typical diode-like characteristic of the device. The inset plot is the gate current profile with $V_g$.* ***c.*** *The ideality factor ($\eta$) extracted at 0 V of $V_g$ as a fit coefficient from I-V characteristic by Shockley diode equation:*

$$I_d = I_S [e^{\frac{V_d + I_S R_S}{\eta V_t}} - 1]$$

*Where, $V_t$=thermal voltage= $KT/e$= 0.025 V, T is the absolute temperature of the p–n junction, $I_S$=reverse saturation current, $\eta$ =ideality factor, $R_S$=series resistance.*

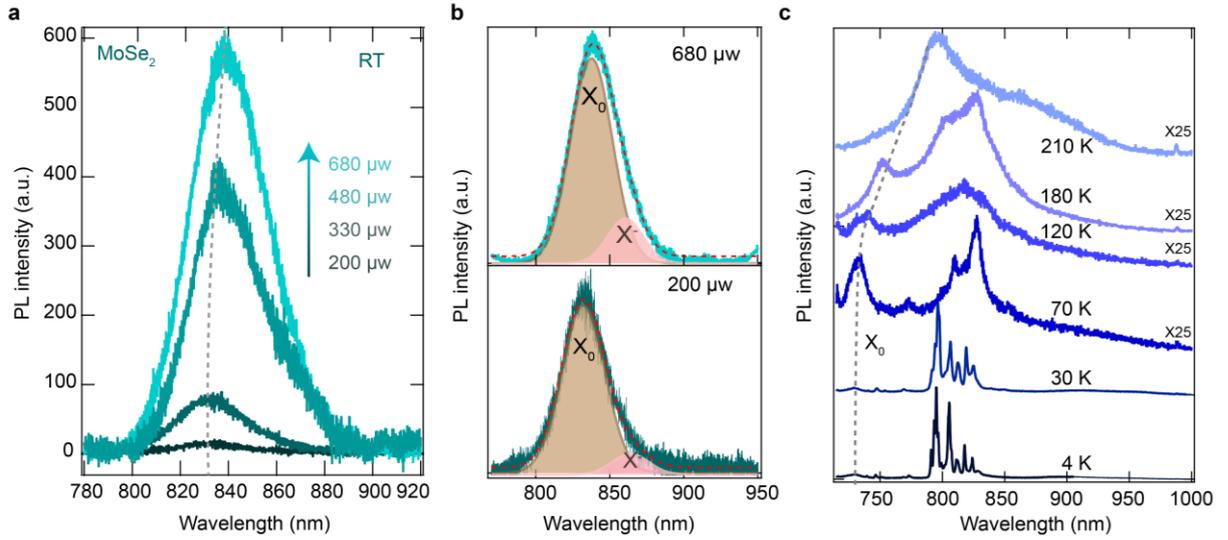

***Figure S4. a.*** *Room temperature PL spectra at MoSe$_2$ for different illumination power.* ***b.*** *The PL peak fit for the neutral exciton and trion at minimum and maximum powers, showcasing the evolution of trion intensity with increasing power, even at room temperature.* ***c.*** *Temperature-dependent PL spectra at WSe$_2$ indicate a redshift of exciton and trap state emissions.*

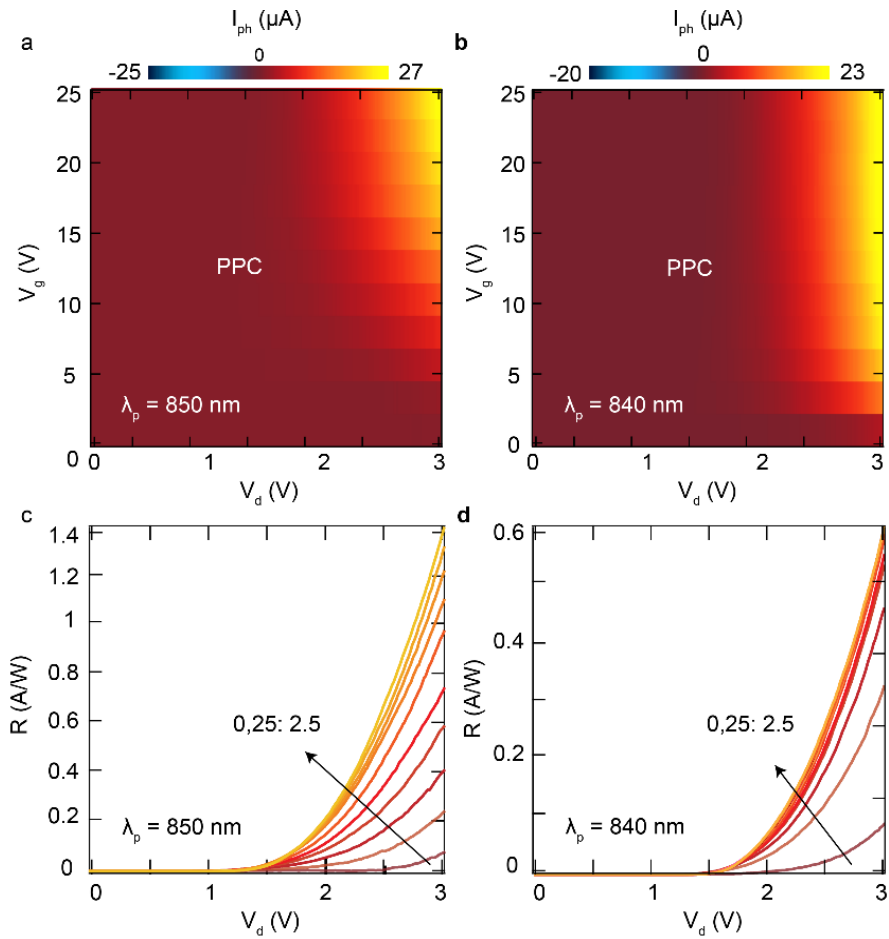

***Figure S5. a,b.*** *Contour plots of $I_{ph}$ for 850 and 840 nm illumination when $V_d$ and $V_g$ are varying from 0 to 3 and 0 to 25 V (interval of 2.5 V), respectively.* ***c,d.*** *Corresponding photoresponse characteristics.*

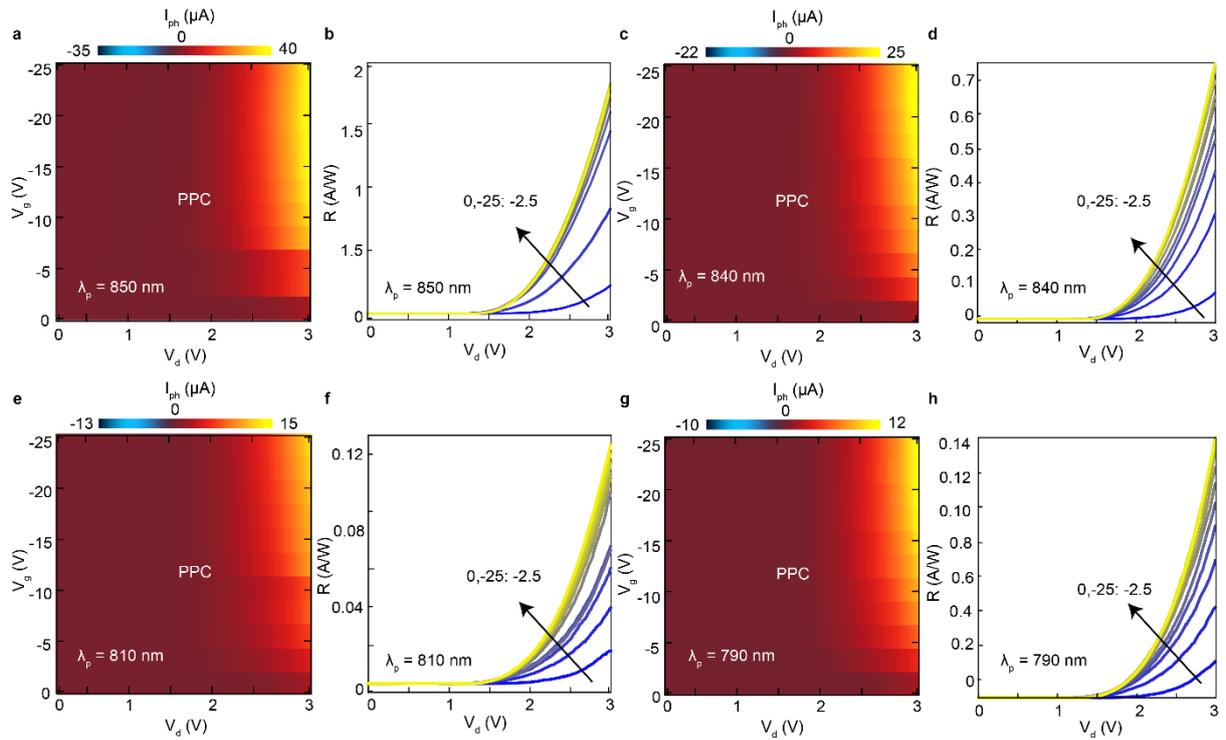

*Figure S6.* Contour plots of $I_{ph}$ and corresponding photoresponse characteristics for *a,b.* 850 nm, *c,d.* 840nm, *e,f.* 810 nm and *g,h.* 790 nm illumination when $V_d$ is varying from 0 to 3 and $V_g$ is varying from 0 to -25 V at the interval of -2.5 V.

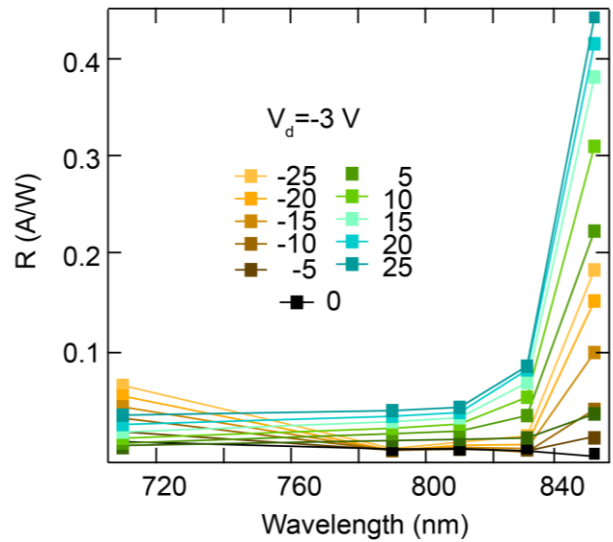

*Figure S7.* The wavelength-dependent responsivity of the n-p-n junction is shown at $V_d$=-3 V.

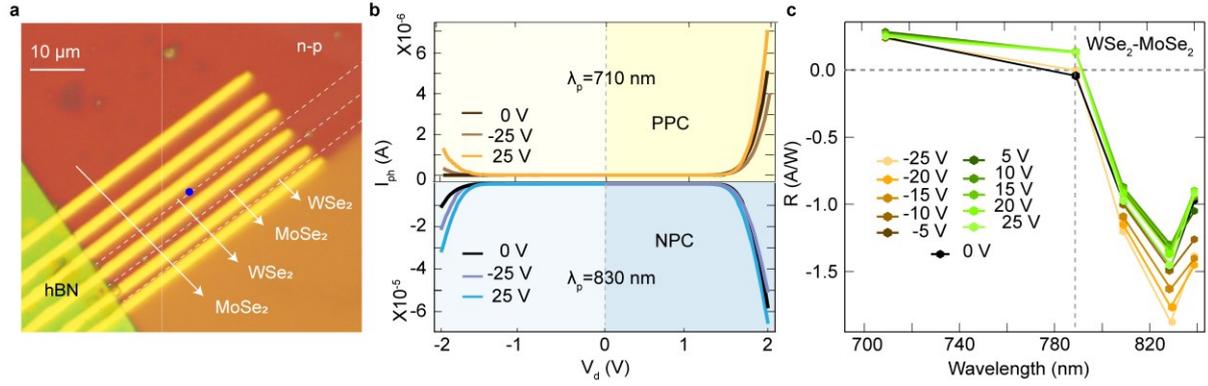

*Figure S8*. **a.** *An optical image of a bilayer one-junction MoSe$_2$-WSe$_2$ LHS device.* **b.** *Photocurrent output characteristics and for the 1-junction p-n device at three different V$_g$ showing PPC at 710 nm and NPC at 830 nm excitation wavelength.* **c.** *The corresponding wavelength-dependent photoresponsivity at different V$_g$ show the opposite NPC-PPC trend in contrast to the n-p-n device configuration.*

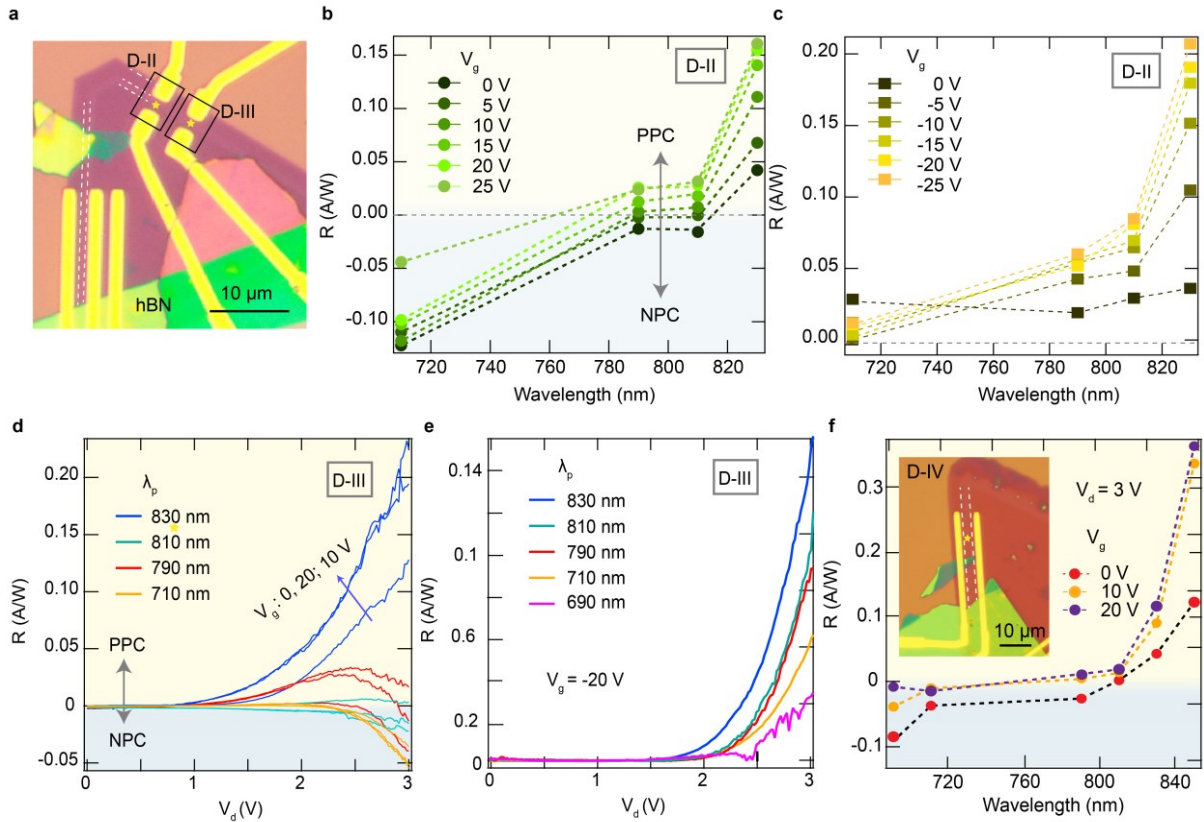

*Figure S9*. *A reproducible NPC-PPC Phenomenon for another two device arrangements with Ti/Au metal contacts.* **a,d.** *Optical images of the n-p-n devices. Tuneable photoresponsivity is shown across the λ$_p$ and +V$_g$, where for -V$_g$ (V$_d$ = 3), R completely shifts to PPC for devices* **b,c.** *D-II and* **d,e.** *D-III.* **f.** *NPC-PPC phenomena for another device arrangement (D-IV) featuring Ti/Pt/Au metal contact; V$_g$ varies from 0 to 20 V at 10 V intervals. The inset shows the optical image of the device.*

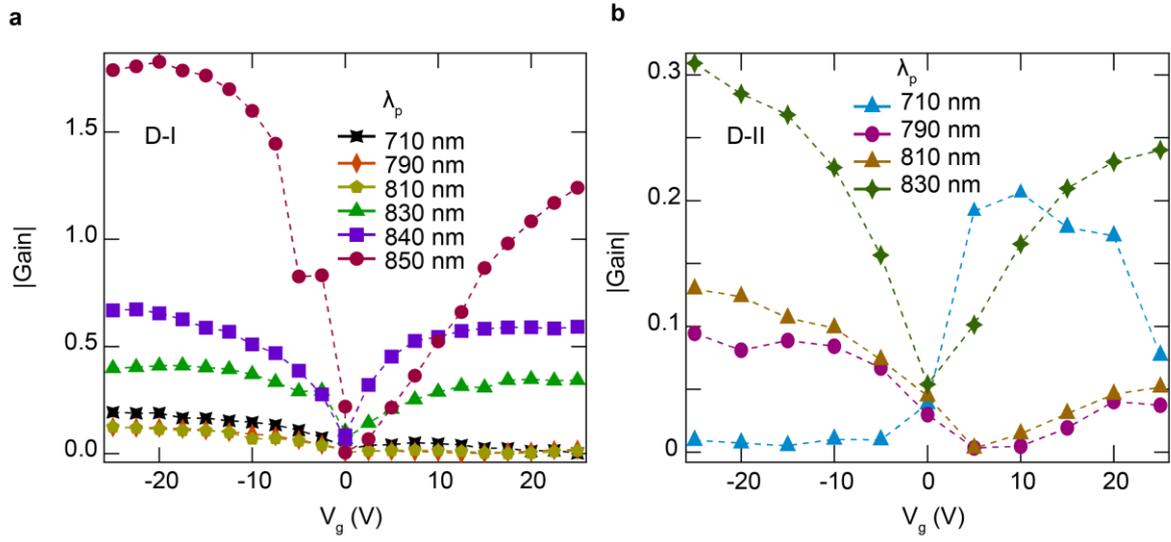

*Figure S10. a,b.* *Device characteristics with absolute photogain vs $V_g$ under various $\lambda_p$ for the two devices, D-I and D-II.*